# A study on Deep Convolutional Neural Networks, Transfer Learning and Ensemble Model for Breast Cancer Detection

Dr. Md Taimur Ahad
Associate Professor
Department of Computer Science and Engineering
Daffodil International Univarsity
Dhaka, Bangladesh
taimur.cse0396.c@diu.edu.bd

Sumaya Mustofa
Department of Computer Science and Engineering
Daffodil International University, Dhaka, Bangladesh
sumaya15-3445@diu.edu.bd

Faruk Ahmed
Department of Computer Science and Engineering
Daffodil International University, Dhaka, Bangladesh
faruk15-4205@diu.edu.bd

Yousuf Rayhan Emon
Department of Computer Science and Engineering
Daffodil International University, Dhaka, Bangladesh
yousuf15-3220@diu.edu.bd

Aunirudra Dey Anu
Department of Computer Science and Engineering
Daffodil International University, Dhaka, Bangladesh
anu15-4170@diu.edu.bd

## Abstract

In deep learning, transfer learning and ensemble models have shown promise in improving computer-aided disease diagnosis. However, applying the transfer learning and ensemble model is still relatively limited. Moreover, the ensemble model's development is ad-hoc, overlooks redundant layers, and suffers from imbalanced datasets and inadequate augmentation. Lastly, significant Deep Convolutional Neural Networks (D-CNNs) have been introduced to detect and classify breast cancer. Still, very few comparative studies were conducted to investigate the accuracy and efficiency of existing CNN architectures. Realising the gaps, this study compares the performance of D-CNN, which includes the original CNN, transfer learning, and an ensemble model, in detecting breast cancer. The comparison study of this paper consists of comparison using six CNN-based deep learning architectures (SE-ResNet152, MobileNetV2, VGG19, ResNet18, InceptionV3, and DenseNet-121), a transfer learning, and an ensemble model on breast cancer detection. Among the comparison of these models, the ensemble model provides the highest detection and classification accuracy of 99.94% for breast cancer detection and classification. However, this study also provides a negative result in the case of transfer learning, as the transfer learning did not increase the accuracy of the original SE-ResNet152, MobileNetV2, VGG19, ResNet18, InceptionV3, and DenseNet-121 model. The high accuracy in detecting and categorising breast cancer detection using CNN suggests that the CNN model is promising in breast cancer disease detection. This research is significant in biomedical engineering, computer-aided disease diagnosis, and ML-based disease detection.

## 1. Introduction

In computer-aided diagnosis (CAD), the Deep Convolutional neural network (D-CNN) has contributed significantly to cancer detection, classification, and segmentation. D-CNN is a collection of techniques that may provide superior results than shallow network architecture for detecting and segmenting the tumorous portion inside a breast Palmer et al [1]. The multi-layered, hierarchical, and block structure of D-CNN can extract low, mid, and high-level characteristics of breast cancer images. In contrast, manual detection and classification of breast cancer require a large quantity of time and effort by doctors and radiologists. Due to the large amount of data generated by scan centers, D-CNN demonstrates exceptional segmentation performance and addresses classification issues within less time and effort, particularly in tasks like the identification of breast cancer cells in Medical Image scans (Sung et al. [2]; Sinthuja et al. [3]).

Machine learning has proven very effective in biomedical engineering (Paul et al. [31]; Shefat et al. [32]. However, two techniques, transfer learning and ensemble learning, have received significant attention in CAD. Transfer learning is a weighted pre-trained CNN version that has been trained on a vast dataset. Using a pre-trained CNN version reduces the trouble of training a CNN from scratch, requiring a large and categorised dataset and several computing powers Han et al. [4] fine-tuning and using CNN as a function extractor transfer learning. During fine-tuning, the weights of the pre-trained CNN models are preserved on specific layers. The preserved layers generally maintain their weight because the capabilities received from those layers are time-consuming, applicable to many tasks, and can be customised during the experiment with different types of datasets Zou et al. [5].

Another method, Ensemble, incorporates many classifiers and has been found to outperform single-classifier methods in terms of accuracy Sagi et al. [6]. Well-known ensemble methods include boosting, bagging, and stacking. The ensemble technique allows a slow-learner algorithm to make final predictions by combining the outputs of a set of basic models Chugh et al.[7]. Generally, the final prediction is made by reducing a loss function based on the cross-validated production of the models to find the base models' optimal weights. The ultimate goal of an ensemble is to correct (compensate) a single model's flaws by combining multiple models, resulting in an ensemble result (prediction and classification) that is superior to any single participating model Sagi et al. [6].

Despite significant research studies devoted to segmenting and detecting cancer detection, there are still knowledge gaps in the literature. Sharma et al. [8] criticised that ensemble models have shown promise in improving breast cancer classification accuracy, but the application of ensemble model breast cancer research is still relatively limited. The researcher purported that further ensemble techniques should be explored in breast cancer detection. There is criticism that imbalanced datasets, where the number of instances belonging to one class (e.g., breast cancer cases) is significantly smaller than the other class (e.g., non-cancer cases), can cause poor performance of the algorithm due to its biases. Chanda et al. [9] reveal that the ensemble model is promising, but the development of current ad-hoc developments overlooks redundant layers and suffers from imbalanced datasets and inadequate augmentation. Accuracy in cancer detection and the classification of modalities are a concern, as lower detection accuracy and large false-positive values will narrow the applicability and acceptability of D-CNN in brain tumor research (Hossain et al. [10]; Aladhadh et al. [11]). Despite data scientists trying to utilise D-CNN, it is still unclear how existing D-CNN architectures perform in detecting cancers. This is because the most prominent CNN architectures, such as VGG, DenseNet, ResNet, and Xception, were tested on small datasets Mohan et al.[12].

Following the gaps, this study aims to detect breast cancer using six original convolutional neural network (CNN) architectures: Inceptionv3, Mobile-NetV2, ResNet18, SE-ResNet152, DenseNet201, and VGG-19, to use transfer learning approach on SEResNet152, MobileNetV2, VGG19, ResNet18, InceptionV3, and DenseNet-121 to see if transfer learning can improve accuracy; and lastly, to develop a hybrid ensembles model called DIR (Densenet121, InceptionV3, ResNet18) aiming to increase the breast cancer detection accuracy. However, Following are the primary contributions and novelties of this study:

1. This study compared the performance of six solitary CNN networks, SE-ResNet152, MobileNetV2, VGG19, ResNet18, InceptionV3, and DenseNet-121, as well as a transfer learning and ensemble model when analysing images for breast cancer detection.
2. A novel 'DIR' ensemble approach is introduced to remove the classification limitations of a singular CNN network. Experiment with three CNN models (Densenet121, InceptionV3, ResNet18) using a weighted voting-based ensemble approach. Multiple comparisons indicate that the DIR ensemble approach provides greater precision in this experiment.
3. The DIR ensemble model compared with the ordinary CNN model and the ensemble model achieved 99.94% accuracy, proving that the ensemble model developed in this article performs better in detecting and classifying breast cancer.

## 2. Related works

Zhang et al. [13] experimented with two different algorithms for detecting and classifying Breast Cancer on Magnetic Resonance Imaging (MRI). Detection was performed using Mask R-CNN, and classification was conducted using ResNet50. The methodology achieved 96% and 81% sensitivity on two different datasets. Another ResNet-50-based breast cancer classification model was developed by Haija et al.[14]. The presented model is based on transfer learning and achieved an accuracy of 99%. This experiment obtained a significant accuracy, but the novelty of the methodology is missing. Yu et al. [15] suggested a customised ResNet-SCDA-50 model where a new data augmentation framework called Scaling and Contrast limited adaptive histogram equalisation Data Augmentation (SCDA) was developed. The proposed model obtained 95.74% of accuracy in classifying breast abnormality. Mahoro et al. [16] proposed a deep CNN and transformer model for breast cancer classification. The authors used TransUNet to segment the breast region and applied four different models where ResNet50 performed best, with an accuracy of 97.26%. Yang et al. [17] developed a model based on CNN and PyQt5, in which improved VGGNet and improved MobileNetV2 algorithms performed classification on two breast cancer datasets. Developing a human-computer interaction GUI system using the PyQt5 library makes this research unique to other existing studies. However, this study didn't mention the accuracy precisely. Mohammed et al. [18] suggested a customised CNN model for feature extraction where the 'Flatten Threshold Swish' (FTS) activation function is used to handle the 'dead neuron' problem. Then, the YOLO loss function was enhanced to effectively handle mammogram lesion scale variation. The proposed methodology achieved 98.72% accuracy for breast cancer diagnosis with only 11.33 million parameters. While other researchers concentrated on deep learning approaches, Jiang et al. [19] concentrated on a bio-inspired Cat Swarm Optimization-guided Convolutional Neural Network (CSO-CNN) algorithm. The suggested algorithm obtained an accuracy of 92.85%. The authors stated that this algorithm is also a mobile network-driven model. Another Swarm Optimization technique was experimented with by Aguerchi et al .[20]. The authors combined Particle Swarm Optimization (PSO) with CNN architecture and proposed a customised Mammography Breast Cancer Classification model. The proposed method obtained 98.23% and 97.98% accuracy on two different datasets. Besides CNN and the variants, few researchers experimented with vision transformers to detect breast cancer. Shiri et al. [21] developed a Supervised Contrastive Vision Transformer (SupCon-ViT) with the inherent strengths and advantages of transfer learning, pre-trained vision transformer, and supervised contrastive learning, resilient to minimal labelled data. The suggested model achieved an F1-score of 0.8188, which is comparatively lower than other existing studies. Ayana et al. [22] proposed a customised vision transformer model with a localisation module for weakly localising critical image features using spatial transformers, an attention module for global learning via vision transformers, and a loss module to determine proximity to a Human Epidermal Growth Factor Receptor 2 (HER2) expression level based on input images by calculating ordinal loss. The proposed model achieved 95% accuracy. Nayak et al. [23] developed a residual deformable attention-based transformer network for breast cancer classification. This customised model used multi-head deformable self-attention mechanisms (MDSA) and residual connections on vision transformer architecture. This model achieved the highest image-level accuracy of 99% and patient-level accuracy of 96.41%.

## 3. Description of experimental method

This research used Google CoLab and the Keras Library to conduct its tests. TensorFlow was utilised since it is one of the top Python deep learning libraries for working with machine learning algorithms on Python. Each model was trained in the cloud with a Tesla GPU and made available through Google's Collaborator platform. For research purposes, the Collaborator framework enables up to 12 GB RAM and a 360 GB GPU in the cloud.

### 3.1 Datasets

The dataset for the study was collected from a public repository. The dataset had two classes: benign and malignant. The images were captured using a microscope and stored as PNG files in RGB format. Figure 1 displays samples of images used in the study.

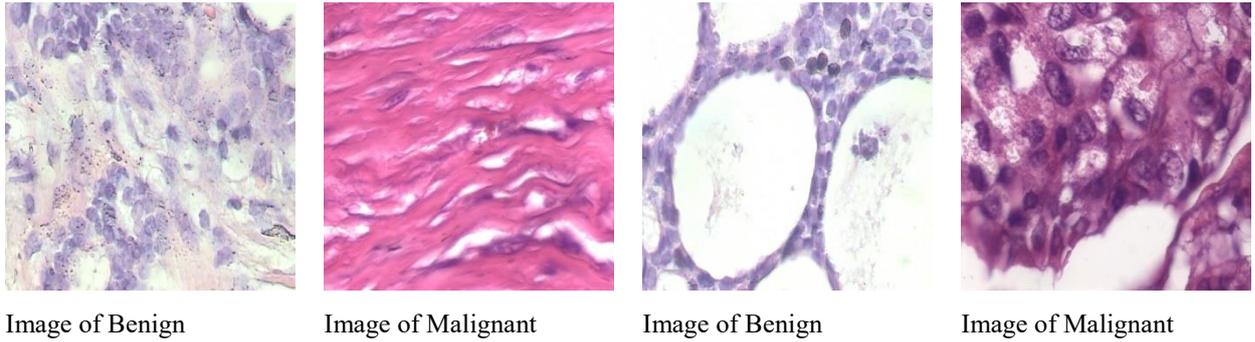

| Image of Benign | Image of Malignant | Image of Benign | Image of Malignant |

Figure 1: Samples of images used in the study

## 3.2 Process of Experiments

The processes of the experiments are described in Figure 3.

**Image Acquisition and Pre-processing**

In this step, the downloaded images from the targeted sites were checked manually to identify if those had a white background. In the case where images (mainly from the BRRI) had colored backgrounds, images were placed on a white background. If disease symptoms such as spots, diseased color, and diseased shape were not visible in an image, that image was removed from the dataset.

**Image Augmentation**

This study used position augmentation such as scaling, cropping, flipping, and rotation, and colour augmentation such as brightness, contrast, and saturation was deployed. Random rotation from −15 to 15 degrees, rotations of multiple of 90 degrees at random, random distortion, shear transformation, vertical flip, horizontal flip, skewing, and intensity transformation were also used in the data augmentation process. This way, ten augmented images from every original image have been created. Figure 2 displays the samples of data augmentations of the images used in the study.

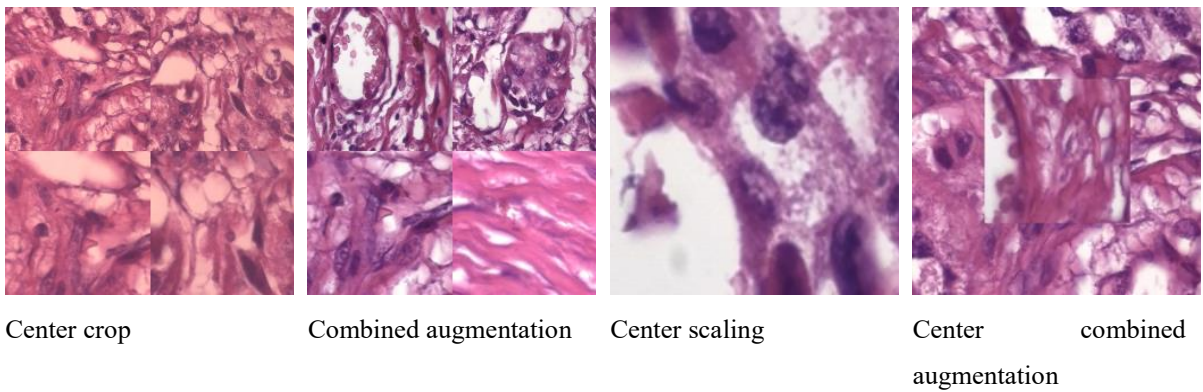

| Center crop | Combined augmentation | Center scaling | Center combined augmentation |

Figure 2: Samples of data augmentations of the images used in the study.

**Experimental Setup**

Table 1. Experimental setup of the experimented algorithms.

| Parameters | Value |
| --- | --- |
| Epoch | 175 (patients = 10) |
| Optimisers | Adam |
| Learning rate | 0.0001 |
| Activation function | ReLU, softmax (last layer) |
| Entropy | Categorical cross-entropy |
| Batch Size | 0.0001 |

**Classification**

In this step, SE-ResNet152, MobileNetV2, VGG19, ResNet18, InceptionV3, and DenseNet-121 were used to automatically detect breast cancer. The neural network was chosen as a classification tool due to its well-known technique of being a successful classifier for many real applications. After the training model, the evaluation model was built for breast cancer detection based on the highest probability of occurrence. Then, the images were classified into different disease classes using a softmax output layer.

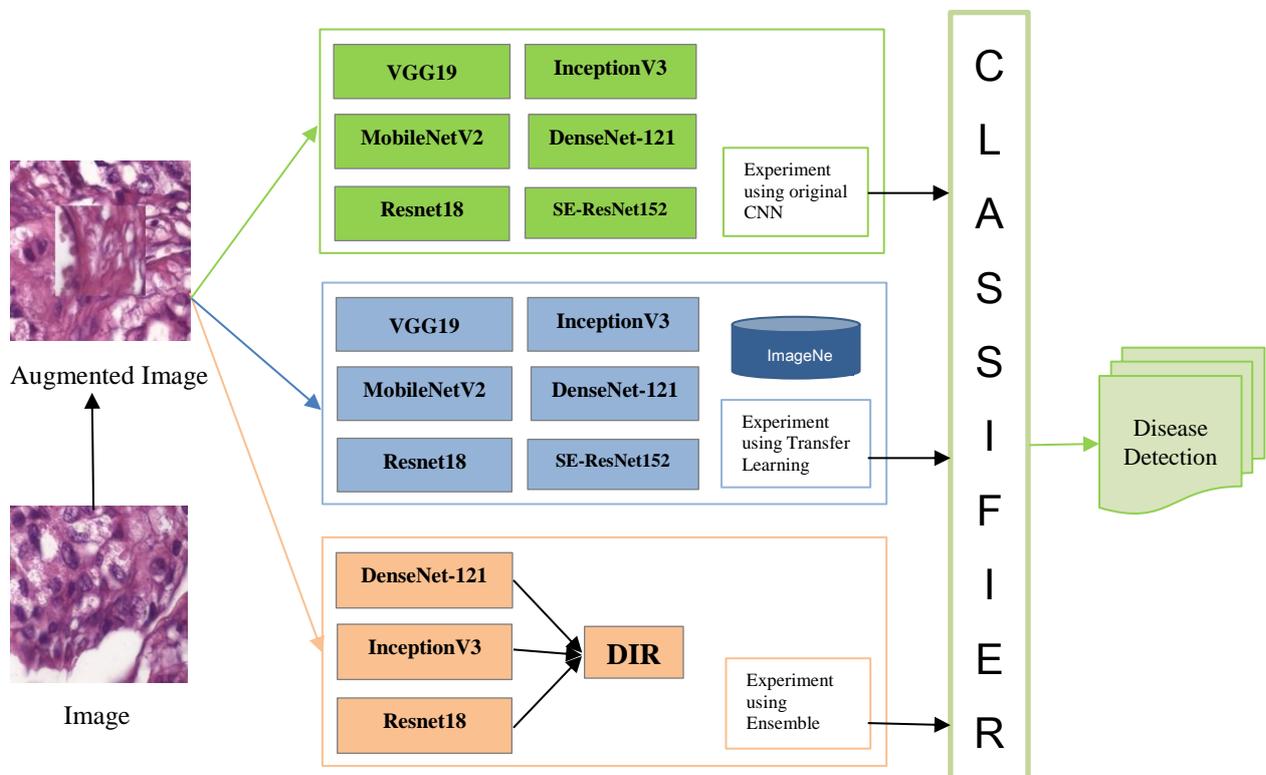

Figure 3: Diagram of the Experiment.

# 4. Results of experiments

## 4.1 Performance metrics

The results of the experiments are measured using the following machine learning classification model performance metrics:

$$\text{Accuracy} = \frac{TP+TN}{TP+TN+FP+FN} \quad (1)$$

$$\text{Precision} = \frac{TP}{TP+FP} \quad (2)$$

$$\text{Recall} = \frac{TP}{TP+FN} \quad (3)$$

$$\text{F1 Score} = 2 * \frac{Precision*Recall}{Precision+Recall} \quad (4)$$

Data loss curves and confusion matrices have also been used to measure the performance of the models.

## 4.2 Experiment 1- Performance of Original CNN network

The performances of the six original individual CNN networks SE-ResNet152, MobileNetV2, VGG19, ResNet18, InceptionV3, and DenseNet-121 are presented in this section. Among them, the DenseNet-121 model had the highest accuracy of 99 %, while the ResNet18 model had the lowest accuracy of 88%.

Table 2. Accuracy for classification of individual CNN networks in detecting breast cancer (original CNN networks).

| Architecture | Training Accuracy | Model Accuracy |
|---|---|---|
| DenseNet121 | 100% | 99% |
| InceptionV3 | 100% | 95% |
| ResNet18 | 100% | 88% |
| SE-ResNet152 | 100% | 94% |
| MobileNetV2 | 100% | 94% |
| VGG19 | 100% | 95% |

Table 3. Precision, recall, f1, and support (n) result of original CNN networks (based on the number of images, n= numbers)

| **DenseNet121** | | |
| --- | --- | --- |
| | **Benign** | **Malignant** |
| Precision | 99% | 99% |
| Recall | 99% | 99% |
| F1-score | 99% | 99% |
| Support (N) | 931 | 989 |
| **ResNet18** | | |
| | **Benign** | **Malignant** |
| Precision | 85% | 90% |
| Recall | 90% | 86% |
| F1-score | 88% | 88% |
| Support (N) | 932 | 988 |
| **SE-ResNet152** | | |
| | **Benign** | **Malignant** |
| Precision | 94% | 93% |
| Recall | 92% | 94% |
| F1-score | 93% | 93% |
| Support (N) | 935 | 985 |
| **InceptionV3** | | |
| | **Benign** | **Malignant** |
| Precision | 94% | 95% |
| Recall | 95% | 95% |
| F1-score | 95% | 95% |
| Support (N) | 926 | 994 |
| **MobileNetV2** | | |
| | **Benign** | **Malignant** |
| Precision | 96% | 92% |
| Recall | 91% | 96% |
| F1-score | 93% | 94% |
| Support (N) | 932 | 98% |
| **VGG19** | | |
| | **Benign** | **Malignant** |
| Precision | 100% | 91% |
| Recall | 90% | 100% |
| F1-score | 95% | 95% |
| Support (N) | 932 | 984 |

Table 3 shows that the precision values for each architecture on the test dataset are considered; the VGG-19, DenseNet-121, and MobileNetV2 architectures provide the best performance. According to the above table, the VGG-19, DenseNet-121, SecrensNet152, and MobileNetV2 models correctly detected and classified breast cancer compared to other models. However, ResNet18 performed poorly, with the lowest identification.

Figure 4 displays the confusion matrix of the original CNNs. Following Table 2, Densenet121 provides a better result, as expected. A total of 919 and 977 images were correctly classified using Densenet201.

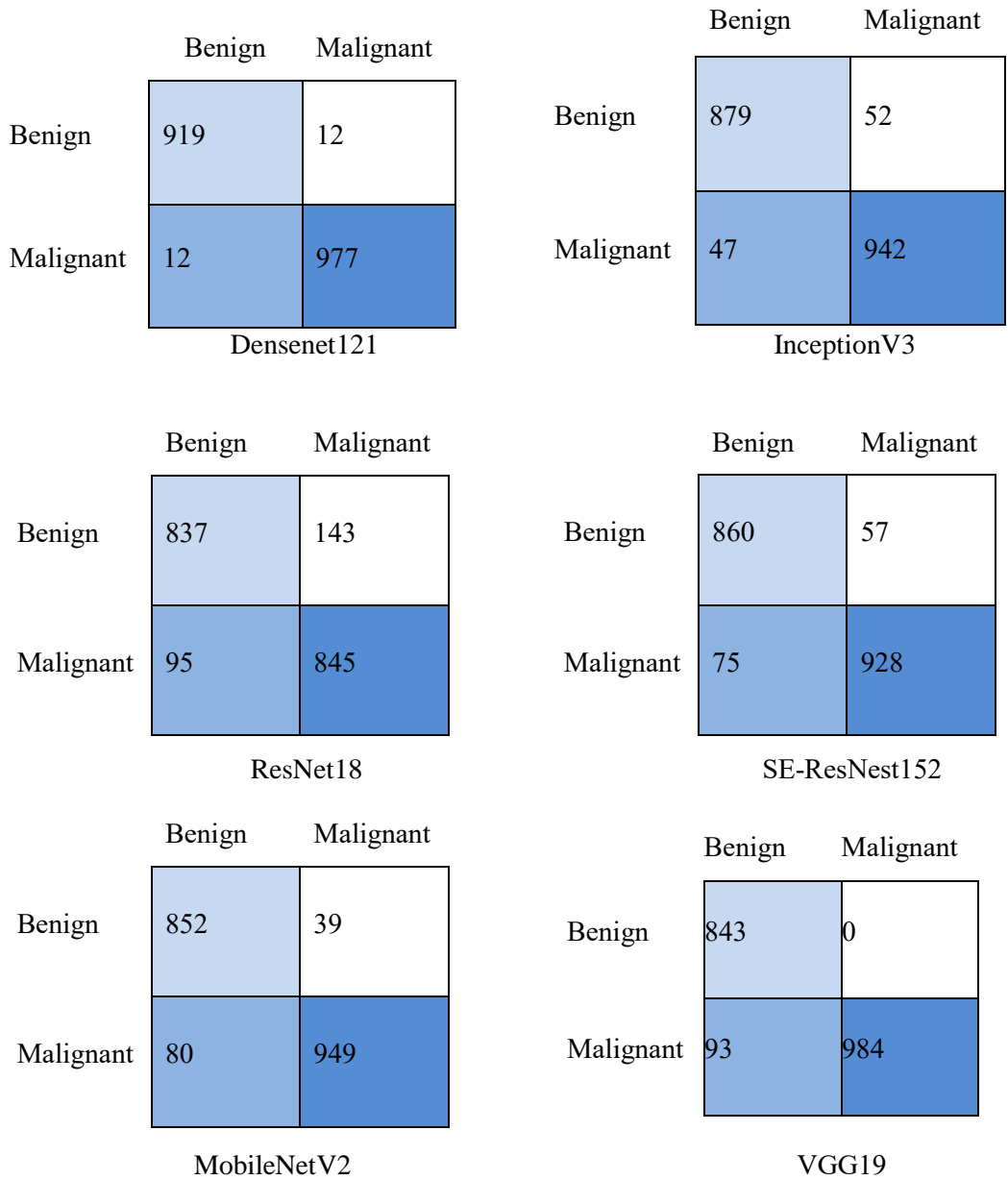

Figure 4: Six confusion matrices of original CNNs.

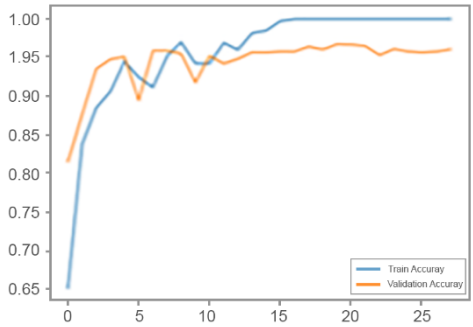
VGG19

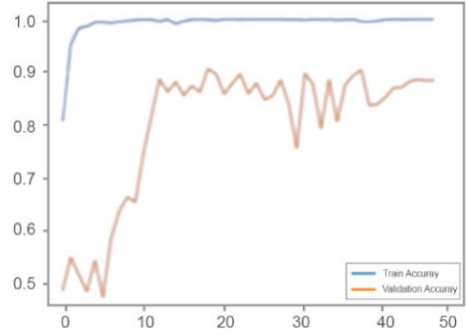
Densenet121

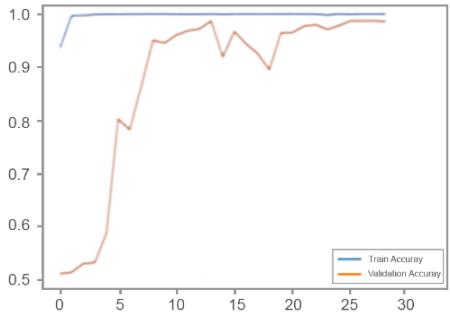
ResNet18

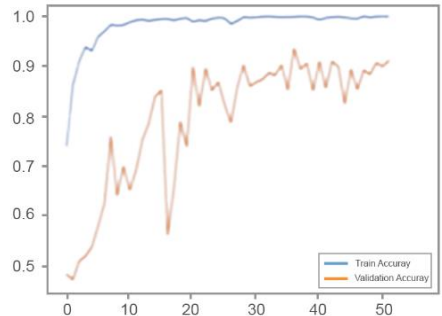
SE-RestNet152

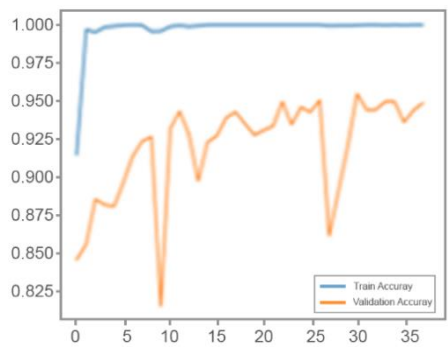
InceptionV3

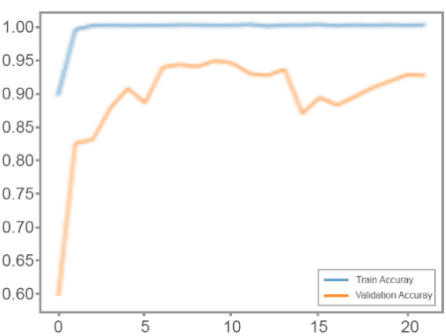
MobileNetV2

Figure 5: The training and validation accuracy of the original CNNs.

Figure 5 depicts the training and validation accuracy of the original model, where the number of epochs is represented on the x-axis, and the accuracy and loss percentages are represented on the y-axis. ResNet18 has the highest, and Densenet121 has the lowest training and validation accuracy over time.

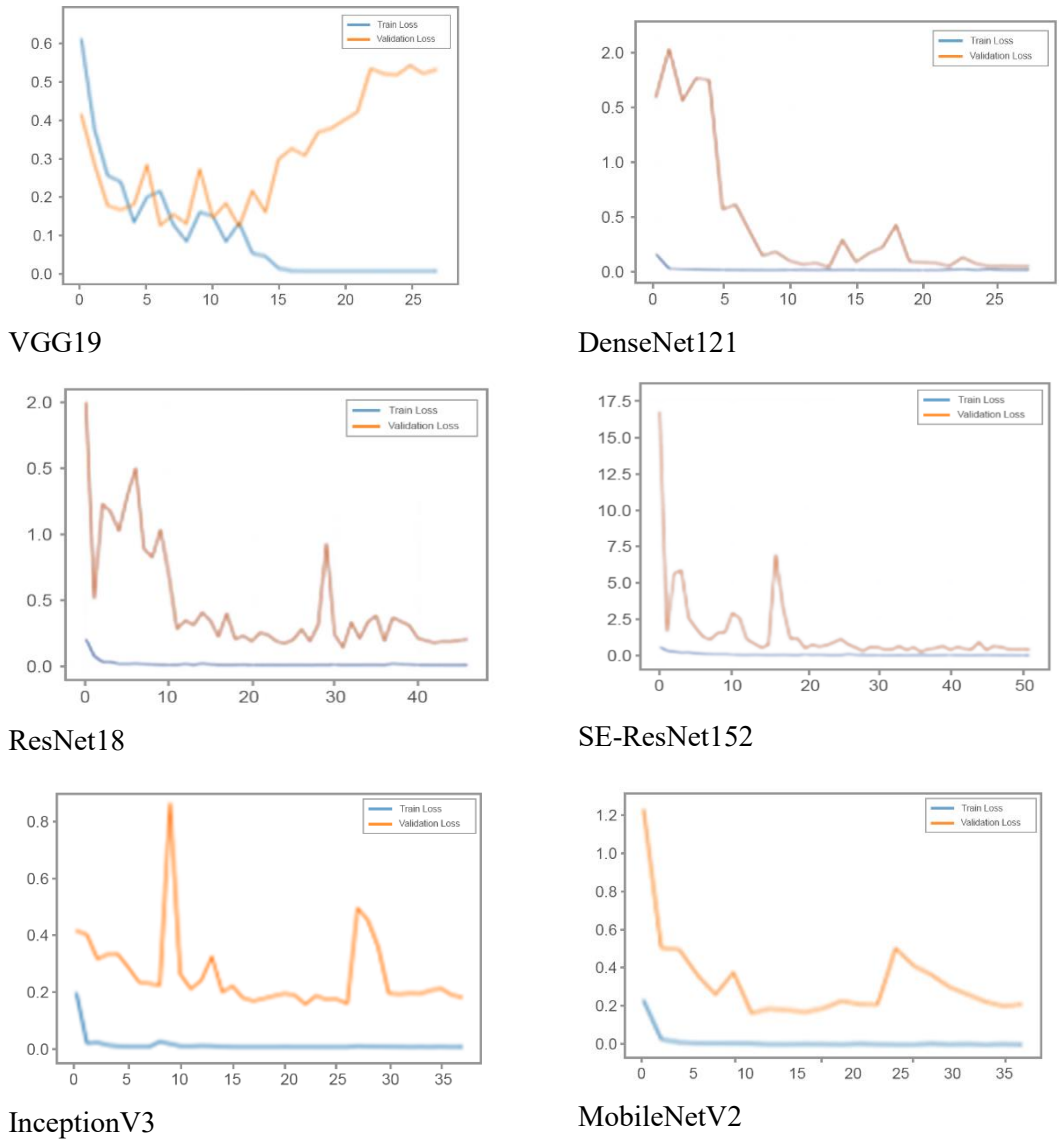

Figure 6: The training and validation losses over the iteration of original CNNs.

Figure 6 depicts the training and validation losses of the original CNN over epochs. Among all the experimented models, DenseNet121 has the lowest training and validation loss, and VGG19 has the highest training and validation loss.

## 4.3 Experiment 2: Transfer Learning CNN Network Accuracy in Detecting Breast Cancer Breast Cancer

Table 4. Transfer learning CNN network accuracy in detecting breast cancer.

| Architecture | Training Accuracy | Model Accuracy |
|---|---|---|
| DenseNet-121 | 95% | 93% |
| InceptionV3 | 95% | 87% |
| MobileNetV2 | 97% | 91% |
| SE-ResNet152 | 97% | 91% |
| ResNet18 | 70% | 64% |
| VGG19 | 81% | 86% |

Six transfer learning CNN architectures' performance is presented in this section. SE-ResNet152, MobileNetV2, VGG19, Resnet18, InceptionV3, and DenseNet-121 models all had high accuracies in the test sets, as shown in Table 4. With an accuracy of 93%, the DenseNet-121 model was the most accurate. The accuracy decreases than the original CNNs' is significant for transfer learning.

Table 5. Precision, recall, f1, and specificity result of CNN networks with transfer learning (n= numbers)

| **DenseNet-121** | | |
|---|---|---|
| | **Benign** | **Malignant** |
| Precision | 98% | 90% |
| Recall | 88% | 98% |
| F1-score | 92% | 94% |
| Support (N) | 932 | 988 |
| **InceptionV3** | | |
| | **Benign** | **Malignant** |
| Precision | 86% | 89% |
| Recall | 78% | 82% |
| F1-score | 95% | 96% |
| Support (N) | 936 | 984 |
| **MobileNetV2** | | |
| | **Benign** | **Malignant** |
| Precision | 95% | 88% |
| Recall | 86% | 96% |
| F1-score | 91% | 92% |
| Support (N) | 934 | 986 |
| **ResNet18** | | |
| | **Benign** | **Malignant** |
| Precision | 93% | 79% |
| Recall | 57% | 97% |
| F1-score | 70% | 87% |
| Support (N) | 942 | 978 |
| **SE-ResNet152** | | |
| | **Benign** | **Malignant** |
| Precision | 95% | 87% |
| Recall | 85% | 96% |
| F1-score | 90% | 91% |
| Support (N) | 946 | 974 |
| **VGG19** | | |
|  | **Benign** | **Malignant** |
| Precision | 91% | 85% |
| Recall | 83% | 92% |
| F1-score | 87% | 88% |
| Support (N) | 932 | 988 |

The Precision, Recall, and F1-score findings from CNN networks incorporating transfer learning are shown in Table 5. Generally, a model with high Precision, Recall, and support is superior. With a 79%, the trial results show that ResNet18 has a low precision in breast cancer breast cancer. DenseNet-121 has the highest precision.

|  | Benign | Malignant |  |  | Benign | Malignant |
|---|---|---|---|---|---|---|
| Benign | 771 | 77 |  | Benign | 820 | 21 |
| Malignant | 161 | 911 |  | Malignant | 112 | 967 |

VGG19                                                                  DenseNet-121

|  | Benign | Malignant |  |  | Benign | Malignant |
|---|---|---|---|---|---|---|
| Benign | 497 | 239 |  | Benign | 802 | 39 |
| Malignant | 445 | 739 |  | Malignant | 144 | 935 |

ResNet18                                                                SE-ResNet152

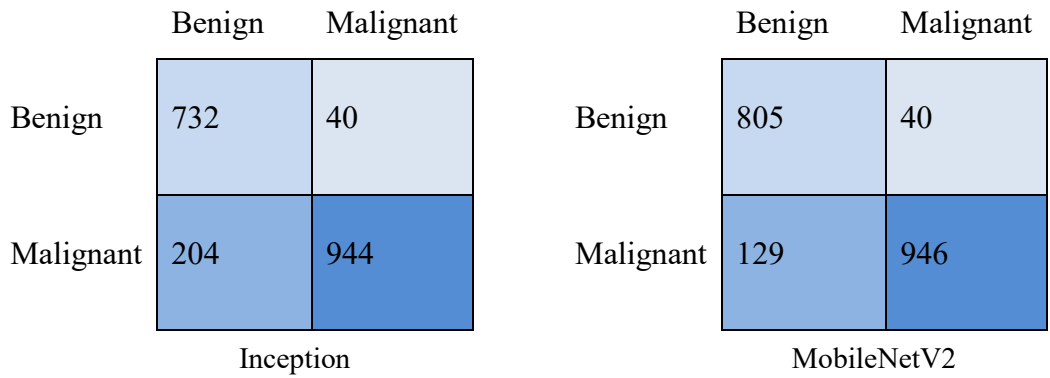

Figure 7: Six confusion matrices of Transfer learning CNNs.

Figure 7 displays the confusion matrix of the original CNNs. Following Table 4, Densenet121 provides a better result, as expected. A total of 820 and 967 images were correctly classified using Densenet201.

Figure 8 shows the Transfer Learning version's training and validation accuracy, with the x-axis representing the number of epochs and the y-axis representing the accuracy and loss chances. ResNet18 has less training and validation accuracy, whereas Densenet121 has the highest training and validation accuracy over time.

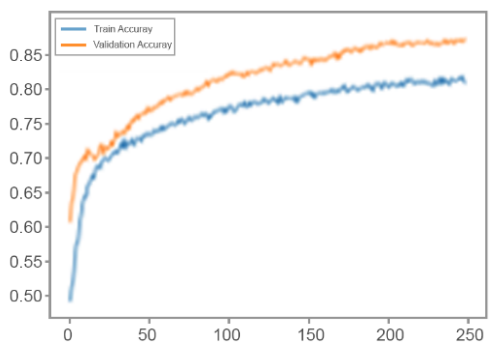
VGG19

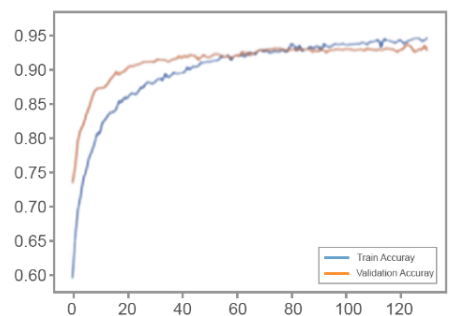
Densenet121

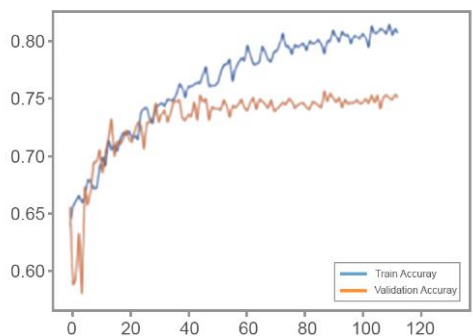
ResNet18

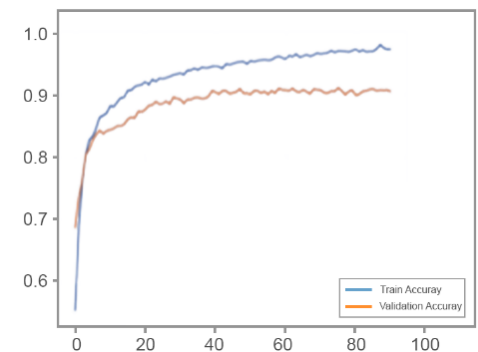
SE-ResNet152

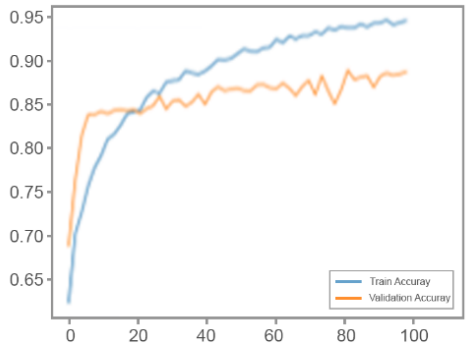
InceptionV3

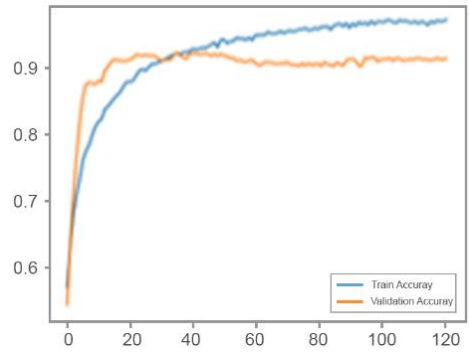
MobileNetV2

Figure 8: The Transfer Learning version's training and validation accuracy.

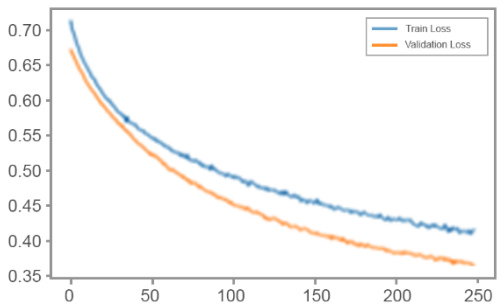
VGG19

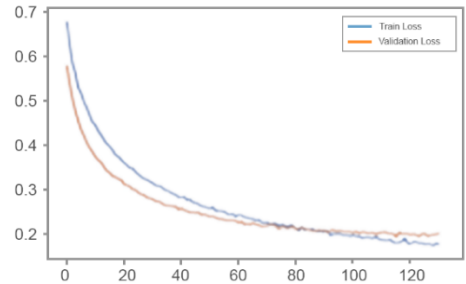
DenseNet121

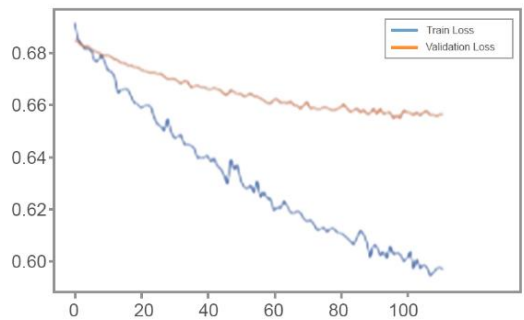
ResNet18

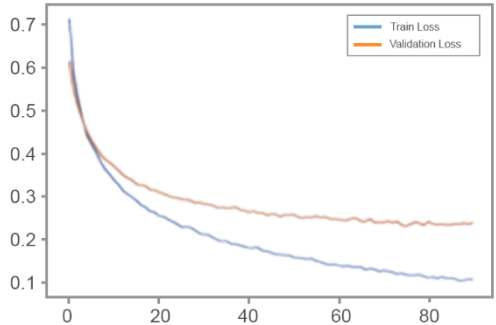
SE-RestNet152

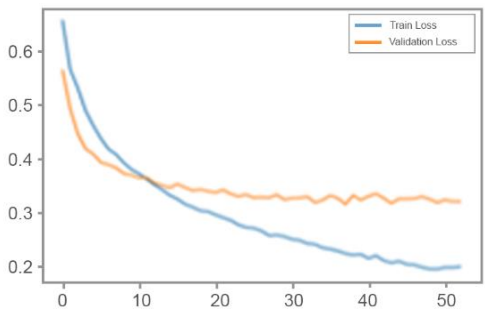
InceptionV3

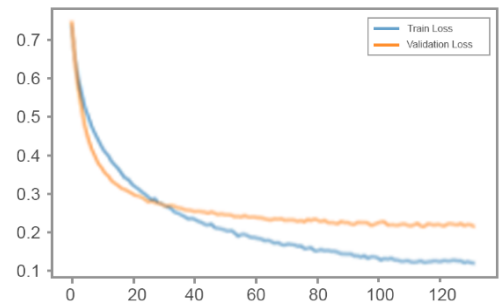
MobileNetV2

Figure 9: Training and validation loss over the iteration.

The training and validation losses of the Transfer Learning are shown in Figure 9 over epochs. CNN uses a loss function to optimise an architecture. ResNet18 has the highest training and validation loss, whereas Densenet121 has the lowest training and validation loss over time.

## 4.4 Experiment 3: Ensemble model development

### 4.4.1 Model development

In this study, one of the main goals was to develop an ensemble model that improves breast cancer detection and classification accuracy. The main reason behind developing the ensemble model is that even if a weak classifier got a wrong prediction, the whole ensemble classifier (robust classifier) could still correct the error. In addition, the ensemble method could reduce the variance. In this experiment, DenseNet121 (99%), InceptionV3 (95%), and ResNet18 (88%) were selected as the candidate for the ensemble model 'DIR' (see Table 2). The purpose was to mix a robust classifier with a weak classifier to validate the ensemble's capabilities, as suggested by Sharma et al.[8].

The ensemble model used in this study aggregates the Sum of Probability values from three CNNs (DenseNet121, InceptionV3, and ResNet18) and calculates the sum of probabilities for each class from the individual CNN architectures. The final prediction is determined by considering the class with the maximum normalised sum.

Our proposed framework uses the ensemble function $\mathcal{E}(\cdot)\ f\ n\ c$ is used, where $n = 3$. Each class receives $n$ confidence values for a given image $I$.

The final classification decision is based on the classes' maximum likelihood. The confidence values $s_{ij}$, where $i \in \{1, 2, \ldots, m\}$ and $j \in \{1, 2, \ldots, n\}$, are aggregated using the Sum of Probabilities.

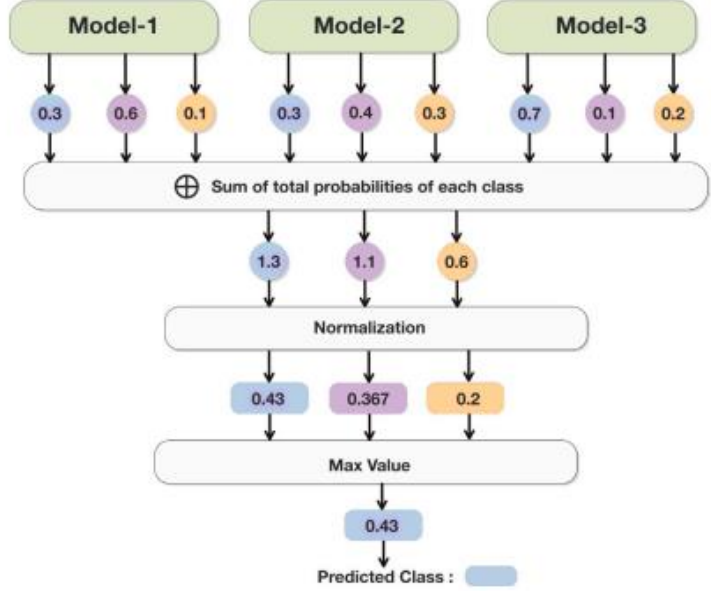

Figure 10. Creation of the proposed DIR ensemble model.

In Figure. 10, the individual summation of prediction values is shown as 1.3, 1.1, and 0.6 for class-1, class-2, and class-3, respectively. The normalisation step ensures that the probability values sum to one, and the class with the most considerable normalised sum is selected as the predicted class. The SoP aggregation process for the proposed network is formulated in Eq. (1), where $i$ is the index for the class, and $j$ is the index for the DCNN models. Thus, $s_{ij}$ Means the prediction value of $i^{th}$ class out of $m$ number of classes and $j^{th}$ model out of $n$ number of models.

A normalisation factor $\sum_{i=1}^{m}\sum_{j=1}^{n} s_{ij}$ is used to normalise the values after the summation of corresponding class values $\sum_{j=1}^{n} s_{ij}$

$$S_{pred} = \max\left(\frac{\sum_{j=1}^{n} s_{ij}}{\sum_{i=1}^{m}\sum_{j=1}^{n} s_{ij}}, \forall i\right) \quad (1)$$

Algorithm 1 Ensemble procedure.

1: Input: [DenseNet121, InceptionV3, ResNet18], test_dataset
2: Output: ensemble_prediction
3: **models** ← [DenseNet121, InceptionV3, ResNet18]
4: for all *model in models* do
5: **predictions** ← model_predict (*test_dataset*)
6: end for
7: **pred_array** ← array (*predictions*)
8: **pred_sum** ← sum (*pred_array, axis* = 0)
9: *ensemble_pred* ← argmax (*pred_sum, axis* = 1)
10: **ensemble_prediction** ← *ensemble_pred*

Table 6. Precision, recall, f1, and specificity result of CNN networks with ensemble techniques (n= numbers)

| **Ensemble Model** ( DenseNet121, InceptionV3, ResNet18) | | |
|---|---|---|
| | **Benign** | **Malignant** |
| Precision | 99% | 95% |
| Recall | 94% | 99% |
| F1-score | 97% | 97% |
| Support (N) | 939 | 981 |

### 4.4.2 Confusion matrix of the ensemble model

The confusion matrix suggests that 892 (TP) Benign and 975 Malignant (TN) cancers were correctly classified. Moreover, the False positives and False negatives were less, impacting the higher model accuracy.

|  | Benign | Malignant |
|---|---|---|
| Benign | 892 | 6 |
| Malignant | 47 | 975 |

Figure 10: Confusion matrix after Ensemble Technique.

Figure 11 and Figure 12 demonstrate a case of a good fit of the ensemble model. The ensemble model data loss curve suggests that achieving a good fit is the goal of the learning algorithm. Training and validation accuracy increased over time while the loss decreased.

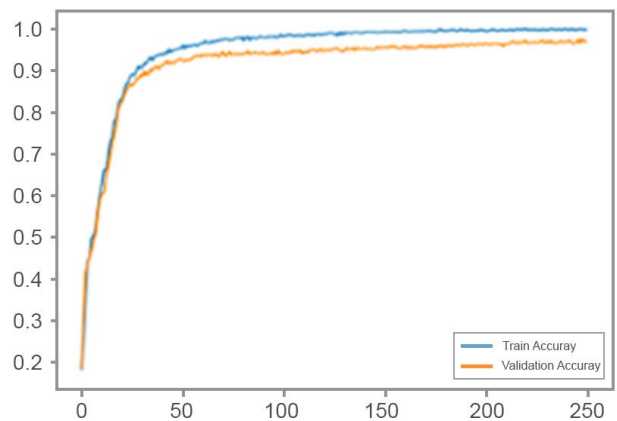

Figure 11: Training and validation accuracy over the epochs of the Ensemble Technique.

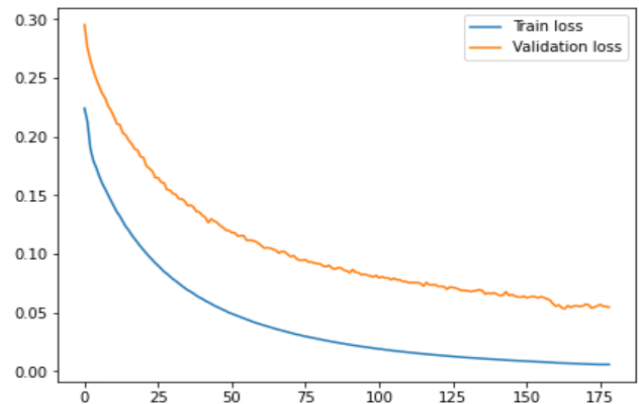

Figure 12: Training and validation loss over the iteration of the Ensemble Technique.

# 5. Discussion

This research presents three (3) different experiments for detecting and classifying breast cancer images using well-known deep learning architectures and identifying the best model that shows promising results.

In the first experiment, a comparison of six CNN networks (DenseNet121, Inceptionv3, ResNet18, SE-ResNet152, MobileNetV2, and VGG19) was conducted on 1920 microscopic breast cancer images of two (2) classes. The experiments suggest that Dense architecture performs best in this experiment. The finding is aligned with Rahman et al. [24], Li et al. [25], and Nayak et al. [23] that Denesenet121 delivers relatively high accuracy. This is because, in DenseNet, each layer obtains a "collective knowledge" from all preceding layers as layers receive inputs from all preceding layers and pass them on to the subsequent layers. The researchers suggested that DenseNet's dense connectivity promotes feature reuse, facilitates gradient flow, and enhances parameter efficiency, improving accuracy in classifying microscopic images. Moreover, DenseNet architecture is superior in capturing fine-grained details and hierarchically representing objects at different scales Li et al. [25].

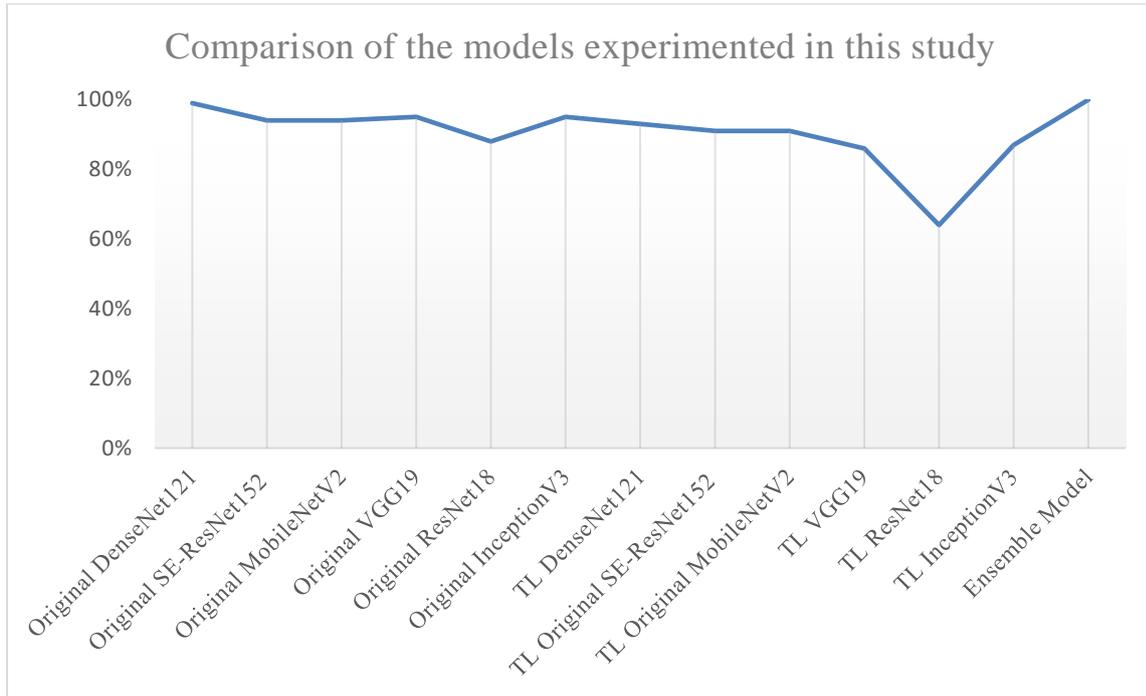

Figure 13: Accuracy comparison among individual CNN, transfer learning, and ensemble models.

The second experiment investigated whether transfer learning, which offers significant advantages in machine learning algorithms, can increase the accuracy of breast cancer detection. However, this study found negative transfer in all six CNN networks, which agrees with the study of Pan et al.[26], Yosinski et al.[27], and Wang et al.[28]. The reason for negative Transfer Learning in Deep Learning is that if the input image differs from the trained data of the ImageNet Dataset, the accuracy will likely decrease. The effect of background noise and the application of different augmentation techniques separately with the test sets resulted in a drop in performance. However, in the case of the original CNN, the model was trained and tested using similar input, and the prediction capabilities were increased in unseen data. Moreover, although the D-CNN can learn features irrespective of the input data, this study's limited number of datasets is likely a factor influencing the prediction capability. Moreover, suppose the pre-trained models were trained on large-scale datasets with natural images, but the target task involves microscopic images or images from a different domain altogether. In that case, the features learned by the pre-trained models may not be relevant or may even be detrimental to the target task. Additionally, negative transfer can occur if the pre-trained models are fine-tuned using a small or insufficient dataset for the target task. Fine-tuning with limited data can lead to overfitting the source task and may not effectively adapt the model to the target task. In this study, the view is also supported by Barbedo et al. [29], who suggested that increasing the dataset size may improve transfer learning performance when the input image is modified using augmentation.

The third experiment aimed to increase the accuracy of breast cancer detection and classification. Not surprisingly, the experiment supports the idea that the effective development of an ensemble model,

'DIR,' can increase the model's accuracy over a single CNN architecture. Our findings also support the study by Jakhar et al.[30], Sharma et al. [8], Jaiswal et al.[31], and Khatun et al.[32].

## 6. Significance of the study

This research provided several significant contributions. This is among the few studies that conducted a comprehensive study on breast cancer detection. Whereas most studies attempted to detect breast cancer using a single CNN or a customised CNN, this study utilised six (6) original CNNs, transfer learning, and ensemble. The promising performance of Dense architecture in detecting microscopic images, negative transfer in cancer images, and the process of developing an ensemble model are some contributions to the body of knowledge. Future data scientists may find valuable information from the technical discussion of the papers.

## 7. Conclusions and future research

Breast Cancer is one of the most common reasons for cancer-associated death. The testing and observations are significant when building models with small cancer datasets. The suggested architecture was compared to transfer learning and six state-of-the-art individual CNN designs in expressions of success. The original and enriched versions of the image dataset were used in the experiments. On both the original and augmented datasets, the Ensemble model outperformed alternative CNN architectures regarding average accuracy and average precision. In this investigation, future studies will need to address several limitations. The study's trials are limited using free resources (Google Colab). Because Google Colab only provides the server for a short time, hyperparameter tuning, basis model training other than ImageNet (this study used ImageNet as the base database), and the usage of Adadelta, FTRL, NAdam, and other optimizers were not included in this research. Another issue is that the study relied on publicly available secondary data rather than original data taken directly from clinical settings or patients.

Checkout and observations are significant when constructing models with small datasets. The ensemble model from DeneNet121, Inceptionv3, and ResNet18 changed and was determined to have the best accuracy in Breast Cancer detection in this study. The recommended architecture was compared to transfer learning and six state-of-the-art individual CNN designs in phases of success. The original and enriched variations of the image dataset had been used in the experiments. On each of the original and more suitable datasets, the ensemble model was determined to be better than current CNN designs regarding average accuracy and precision. With high computation resources and big data, we aim to be able to forecast and provide accurate outcomes in the future. Increase the number of frames per second and boost overall performance with more computational resources. In the future, we hope to develop a user interface for detecting and localising breast cancer. This interface would detect ailments and provide instructions on how to control them. We aim to develop a mobile phone-based breast cancer diagnosis application tool because mobile phones are a favored technological gadget among consumers of developing countries.

## Declaration of competing interest

The authors declare that they have no known competing financial interests or personal relationships that could have appeared to influence the work reported in this paper.

## Data availability

The data of this research are stored in the Kaggle repository.

## Funding Statement

No funding was provided to support the work, and neither did any of the researchers receive funds.